\documentclass[aps,prl,showpacs,superscriptaddress,twocolumn]{revtex4-1}
\usepackage{lmodern}

\usepackage[latin9]{inputenc}
\usepackage{amsmath}
\usepackage{amssymb}
\usepackage{graphicx}
\usepackage{epstopdf}
\usepackage{color}
\usepackage{enumerate}
\usepackage{ulem}

\begin{document}
\bibliographystyle{apsrev4-1}

\title{High field termination of a Cooper-pair insulator}

\author{B. Sac\'{e}p\'{e}}
\author{J. Seidemann}
\affiliation{Univ. Grenoble Alpes, F-38000 Grenoble, France}
\affiliation{CNRS, Institut N\'{e}el, F-38000 Grenoble, France}
\author{M. Ovadia}
\author{I. Tamir}
\author{D. Shahar}
\affiliation{Department of Condensed Matter Physics, Weizmann Institute of Science, Rehovot 76100, Israel}
\author{C. Chapelier}
\affiliation{Univ. Grenoble Alpes, F-38000 Grenoble, France}
\affiliation{CEA, INAC-SPSMS, F-38000 Grenoble, France}
\author{C. Strunk}
\affiliation{Institut f\"ur experimentelle und angewandte Physik, Universit\"at Regensburg, Regensburg, Germany}
\author{B. A. Piot}
\affiliation{Laboratoire National des Champs Magnétiques Intenses (LNCMI), CNRS-UJF-UPS-INSA-EMFL, F-38042 Grenoble, France}

\begin{abstract}
We conducted a systematic study of the disorder dependence of the termination of superconductivity, at high magnetic fields ($B$), of amorphous indium oxide films. Our lower disorder films show conventional behavior where superconductivity is terminated with a transition to a metallic state at a well-defined critical field, $B_\textrm{c2}$. Our higher disorder samples undergo a $B$-induced transition into a strongly insulating state, which terminates at higher $B$'s forming an insulating peak. We demonstrate that the $B$ terminating this peak coincides with $B_\textrm{c2}$ of the lower disorder samples. Additionally we show that, beyond this field, these samples enter a different insulating state in which the magnetic field dependence of the resistance is weak. These results provide crucial evidence for the importance of Cooper-pairing in the insulating peak regime.

\end{abstract}
\maketitle

When a highly disordered, superconducting, film is subjected to a strong magnetic field ($B$) it can undergo a transition to an insulating state \cite{Goldman98,Gantmakher10}. In recent years this insulator drew significant attention because of the prospect that the charge carriers in it are Cooper-pairs. Evidence for this comes from a variety of different materials and experimental techniques, such as low-temperature ($T$) transport \cite{Hebard90,Gantmakher98,Murthy04,Murthy05,Steiner05,Baturina07a,Baturina07b,Steiner08,Allain12,Ovadia13}, measurements of Little-Parks oscillations \cite{Stewart07,*Nguyen09,Kopnov12}, microwave-frequency conductivity measurements \cite{Crane07a,*Crane07b} and tunneling \cite{Sacepe07,Sacepe10,Mondal11,Sacepe11,Chand12,Kamlapure12,Sherman12,Noat13}. Similar indications arrive from several theoretical works that attempt to explain these experimental results \cite{Fisher90,Galitski05,Dubi06,Dubi07,Vinokur08,Muller09,Pokrovsky10,Feigelman10a,Feigelman10b,Bouadim11,Muller13}.

A key ingredient implicit in this picture relates to the high-$B$ limit of this Cooper-pair insulator (CPI). If this insulator is comprised of (localized) Cooper-pairs one expects that, at high enough $B$ such that the pairs are locally broken \cite{Gantmakher98}, the CPI will cease to exist and a $B$-induced transition to a different phase will take place. In this Letter we experimentally study this transition in thin-films of amorphous indium oxide (a:InO) and show that this higher $B$ phase is also an insulator, distinct from the CPI, and that the transition between the two insulators takes place at a $B$-range that is close to the microscopic critical field for superconductivity, $B_\textrm{c2}$. The coincidence of $B_\textrm{c2}$ with the rapid termination of the insulating peak is a central indication for the role of Cooper pairs in the formation of the CPI.

\begin{figure}[tb!]
\includegraphics[width=\linewidth,bb=72 32 618 886]{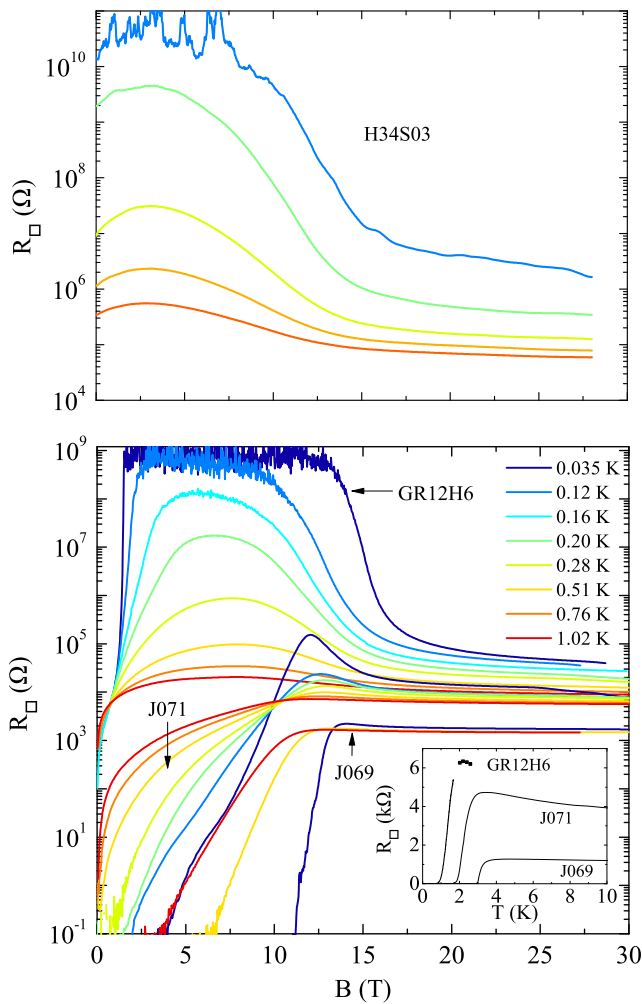} 
\caption{\label{Fig1} $R_{\square}$ vs. $ B $ obtained from four a:InO films with different disorder, measured at $T$ ranging from 0.035 K to 1 K.  The highest disorder sample, H34S03, is insulating at $B=0$ (top). The lowest disordered samples, J069 and J071, were measured in a 4-terminal configuration with AC excitation current of 10 nA. The higher disordered films, GR12H6 and H34S03 were measured in a 2-terminal configuration with AC excitation voltages of 100 $ \mu $V and 500 $ \mu $V, respectively. The inset shows the $B=0$ superconducting transitions of the three superconducting films.}
\end{figure}

To achieve the results of this work we conducted a comparative study of a wide range of disorder levels in a:InO films. The samples were prepared on a SiO$_2$ substrate cleaned and plasma polished immediately prior to the e-gun deposition using 99.99\% In$_2$O$_3$ pellets in a high-vacuum chamber with a controlled O$_2$ partial pressure of $1.5-15\cdot 10^{-5}$ Torr. The samples, ranging in size from 20-200 $\mu$m, were contacted using pre-deposited Au-Ti pads. The measurements were carried out at the high-$B$ facility in Grenoble, using a dilution refrigerator capable of base $T\geq 0.035$ K mounted in the bore of a resistive magnet capable of $B=30$ T. Transport measurements were conducted using low-frequency (typically 3 Hz) 4-terminal lock-in technique for $R<1 \,M\Omega$ and extended to higher $R$ ($<10^{10}\, \Omega$) using 2-terminal AC measurements \cite{Ovadia09}. Wherever the two methods overlapped in range the results were in agreement within experimental noise and error.

The results of these measurements are represented in Fig. \ref{Fig1} where, depending on disorder, several distinct types of behavior can be seen. The lowest disorder samples, such as J069, provide insight into the superconducting properties far from the transition to the insulator. They exhibit a transition into a superconducting state at $T=T_c\approx 3$ K and, when a strong perpendicular $B$ is applied, revert to the normal state with a slight hump observed only at our lowest $T$'s (see sample J069 in Fig. \ref{Fig1}). Since such a behavior belongs to the well-understood realm of conventional dirty superconductors \cite{DegennesBook} we can use the study of the $T$ and disorder dependence of their $B_\textrm{c2}$ to extract parameters that will be relevant for the more disordered samples where they are much harder to evaluate.

\begin{figure}[b!]
\includegraphics[width=1\linewidth]{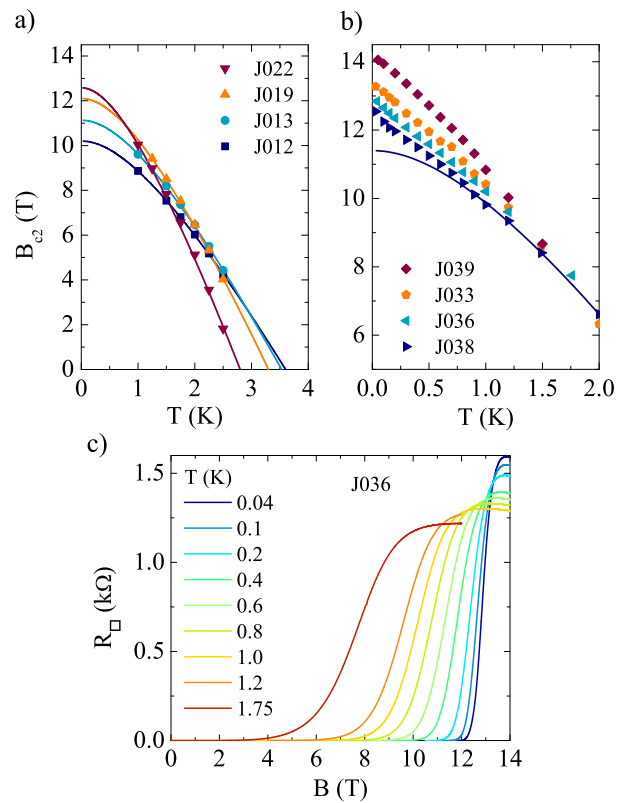}
\caption{\label{Fig2} (a) $B_\textrm{c2}$ vs. $T$ for low disordered a:InO samples (see Table \ref{table1}). 
(b) $B_\textrm{c2}$ vs. $T$ in our low-$T$ range for four different low disordered a:InO samples including J036 (see (c)). The solid lines in (a) and (b) are the BCS-fitting curves \cite{DegennesBook}.(c) $ R_{\square} $ versus $ B $ for one of our cleanest a:InO samples, J036, which is similar to J013 and J069 at $T$ from 0.044 K to 1.75 K. }
\end{figure}

To extract these parameters we plot, in Fig. \ref{Fig2}a, $B_\textrm{c2}$ vs. $T$, near $T_c$, taken from four of our low-disorder samples whose parameters are given in Table \ref{table1}. We define $B_\textrm{c2}$ at the half-point of the transition of the $R$ vs. $B$ curves and verified that other choices of the transition point lead to similar results. 

Inspecting Fig. \ref{Fig2}a we first note that the slope of $B_\textrm{c2}(T)$ in the vicinity of $T_c$ increases with disorder level (indicated by $R_\square^{300K}$). This behavior reflects the reduction of the diffusion coefficient $D$ by disorder that, in turn, increases the slope $\left( dB_\textrm{c2}(T)/dT \right) _{T \rightarrow T_c} = -4\varPhi_0 k_B/ \pi^2 \hbar D $, where $\phi_0 = h/2e$ is the flux quantum \cite{DegennesBook}. For these four samples $D$ is reduced from $0.28\, cm^2/s$ (sample J012) to $0.18\, cm^2/s$ (sample J022).

Another consequence of the increase in disorder is the increase of the extrapolated $B_\textrm{c2}(0)$. Disregarding spin effects $B_\textrm{c2}(0)$ is directly related to the (dirty) superconducting coherence length $\xi_d$ through $B_\textrm{c2} (0) = 0.69 \phi_0/2\pi \xi_{d}^2$ \cite{DegennesBook}. The increase in $B_\textrm{c2}(0)$ corresponds to a decrease of $\xi_d$, and the above relation yields $\xi_d\approx 4.7-4.0\,nm$ for our samples\cite{Steiner05}. These estimates are consistent with those obtained by using the superconducting gap value $\Delta \sim 0.55 \,meV$ measured by tunneling \footnote{See Supplementary Material} and the values of $D$ in Table \ref{table1} yielding 
$\xi_d = 0.83 \sqrt{\hbar D/ \Delta}  \sim 4.7-3.7 \, nm$ \cite{DegennesBook}. 

\begin{table*}[tb!]
\centering
\caption{\label{table1}Parameters of the superconducting films. All films are $60 \,nm$ thick, excepted J071, GR12H6 and H34S03 which are $30 \,nm$ thick. $R_\square^{max}$  is the maximum resistance reached before the superconducting transition; $B_\textrm{c2}^{fit}(0)$ is obtained from the BCS fit of the $B_\textrm{c2}(T)$ curves; $B_\textrm{c2}^{meas}(0.035\,K)$ is the measured value at $T=0.035\,K$.}
\begin{ruledtabular}
\begin{tabular}{llllllll}
$Sample$ & $R_\square^{max} \,(k\Omega)$ & $R_{\square 300K} \,(k\Omega)$ & $T_c \,(K)$ & $B_\textrm{c2}^{fit}(0) \,(T)$ & $B_\textrm{c2}^{meas}(0.035\,K) \,(T)$ & $D \,(\text{cm}^2/\text{s})$ & $\xi_d \,(nm)$ \\ \hline \hline
J012 & 0.5 & 0.4 & 3.5 & 10.2 & - & 0.28 & 4.7 \\
J013 & 0.7 & 0.5 & 3.4 & 11.1 & - & 0.26 & 4.5 \\
J019 & 1.3 & 0.7 & 3.1 & 12.1 & - & 0.22 & 4.3 \\
J022 & 2.3 & 1.0 & 2.7 & 12.6 & - & 0.18 & 4.2 \\
J033 & 1.3 & 0.8 & 3.2 & 12.5 & 13.3 & 0.20 & 4.1 \\
J036 & 1.0 & 0.6 & 3.5 & 11.9 & 12.8 & 0.23 & 4.2 \\
J038 & 1.2 & 0.7 & 3.5 & 11.7 & 12.6 & 0.23 & 4.2 \\
J039 & 1.7 & 0.9 & 3.0 & 13.3 & 14.1 & 0.18 & 4.0 \\
J069 & 1.3 & - & 3.1 & 11.9 & 13.0 & 0.21 & 4.2 \\
\hline
J071 & 4.7 & - & 2.3 & N/A & N/A & N/A & N/A \\
GR12H6 & 6.3 & - & 1.2 & N/A & N/A & N/A & N/A \\
\end{tabular}
\end{ruledtabular}
\end{table*}

For four of our low-disorder samples we measured the magnetoresistance isotherms down to $T=0.035$ K in order to approach $B_\textrm{c2}(0)$ limit. Interestingly the resulting $B_\textrm{c2}(T)$, shown in Fig. \ref{Fig2}b, do not follow the expected behavior of type II superconductors (solid line); instead it decreases linearly with $T$ down to our lowest $T$. This linear decrease of $B_\textrm{c2}(T)$ is accompanied by sharper $B$-driven transitions to the normal state, shown in Fig. \ref{Fig2}c, indicating a smaller contribution from vortex creep at low $T$'s. A similar upturn of $B_\textrm{c2}(T)$ at low $T$ have been observed in similar systems \cite{Hebard84,Furubayashi85}. For disordered superconductors, recent theoretical works accounted for this deviation by including the effects of superconducting inhomogeneities induced by mesoscopic fluctuations \cite{Galitski01}. Inhomogeneities in such systems were observed in scanning tunneling experiments \cite{Sacepe07,Sacepe11,Mondal11,Kamlapure12,Noat13}.

Given our $T_c$ values of 2.7 to 3.5 K, the corresponding $B_\textrm{c2}(0)$'s are unexpected. In conventional type II superconductors, $B_\textrm{c2}(0)$ is determined by orbital effects and rarely exceeds the Pauli paramagnetic limit given by $B_\textrm{p} [T] \simeq 1.8\,T_c [K]$ \cite{Clogston62,Chandrasekhar62}. The fact that $B_\textrm{c2}(0)$ significantly exceeds the Pauli limit in our samples indicates that the effective spin susceptibility is reduced, most likely due to spin-orbit scattering by the indium atoms \cite{Anderson59}.

Because the spin susceptibility is corrected by spin-orbit scattering according to $\chi_S / \chi_N \simeq 1 - 2 \Delta \tau_{so}/\hbar$ for $\Delta \ll h/\tau_{so}$, where  $\chi_{S(N)}$ is the spin susceptibility in the superconducting (normal) state and $\tau_{so}$ the spin-orbit scattering time \cite{Anderson59}, the resulting correction to the Pauli field \footnote{M. Feigel'man private communication}: $B_\textrm{p} \propto \frac{1}{\mu_B} \frac{\Delta } {\sqrt{\Delta\, \tau_{so}/\hbar}}$ leads to an increase of $B_\textrm{p}$ with spin-orbit scattering rate.
In our samples $\tau_{so}$ can be readily estimated from the elastic scattering time $\tau $ through $\tau_{so} \simeq \tau / (Z\alpha)^4 $, where $Z$ is the atomic number of the scattering atom and $\alpha \simeq 1/137 $ the fine-structure constant \cite{Abrikosov62,Rogachev05}. With $\tau \sim 3.10^{-16}\,s$ extracted from the room temperature Boltzmann conductivity of sample J012, one obtains $\tau_{so} \sim 2.10^{-14}\,s$, leading to an enhancement of $B_p$ by a factor  $1 /\sqrt{\Delta \tau_{so}/\hbar} \sim 7$. These rough estimates point to a purely orbital suppression of superconductivity at $B_\textrm{c2}$. We note that, as the superconducting state is three dimensional (thickness $> \xi_d$), we cannot use parallel field magnetoresistance measurements to disentangle the spin from the orbital contributions to $B_\textrm{c2}$.

We now turn to the higher disorder a:InO samples. For these samples $B_\textrm{c2}$ can no longer be identified because it is masked by the appearance of a pronounced magnetoresistance peak. Instead superconductivity is terminated by another transition, at a well-defined (and lower) $B_C$, to a strongly insulating state \cite{Murthy04}. The strength and position of the peak, as well as $B_C$, evolve with disorder as can be seen in the two intermediate-disorder samples in Figure \ref{Fig1}, J071 and GR12H6. For sample J071, $R$ at the peak is $\sim 10^5\,\Omega$ (at $T=0.035\,K$) and the peak's location is at $B=12$ T while for GR12H6 $R$ is beyond our measurement capability at $T<0.15$ K, and the peak is located at $\approx 7.5$ T. For these two samples $B_C=10$ and $0.8$ T respectively.

For the highest disorder sample in this study, H34S03, superconductivity is no longer observed and the sample is insulating at $B=0$. Despite this marked difference a similar insulating peak is found at $B\approx 5$ T.\cite{Baturina07a}

The large range of disorder spanned by the samples in this study enables the central observation of this work. Despite the vastly different low-$B$ characteristics of the four samples in Figure \ref{Fig1}, they all undergo a dramatic change in $R$ at a similar $B$-range of 13-15 T. For our lowest disorder sample, J069, we have already identified this region with the critical $B$ for superconductivity, $B_\textrm{c2}$. The large, sometimes precipitous, drop of $R$ terminating the insulating peak in the higher disorder samples occurs at the same $B$-range.

For less disordered thin-film superconductors, $B_\textrm{c2}$ marks the $B$ where superconductivity is terminated by elimination of Cooper pairing. The coincidence of $B_\textrm{c2}$ with the $B$ that terminates the CPI in the three higher disorder samples implies a remarkable possibility: if we assume that, in the insulating peak, transport is maintained by Cooper-pairs that are localized in space, then breaking these pairs by applying $B>B_\textrm{c2}$ should destroy the CPI and strongly modify the transport in the sample.

If the termination of the insulating peak is a result of full pair-breaking at $B\sim B_\textrm{c2}$ a natural question arises: what is the electronic state at $B>B_\textrm{c2}$ for our higher disorder samples. The high-$B$ state beyond the CPI in titanium-nitride and a:InO has been recently studied by Baturina {\it et al.} \cite{Baturina07a}. They found that its resistance is metallic with an extrapolated $T=0$ value close to the quantum unit, $h/e^{2}$, leading them to propose the existence of a quantum metal phase \cite{Butko01} in these materials at high-$B$ and $T=0$. This intriguing possibility has prompted us to study our films at extremely high-$B$'s and low-$T$'s.

\begin{figure}[tb!]
\includegraphics[width=0.9\linewidth]{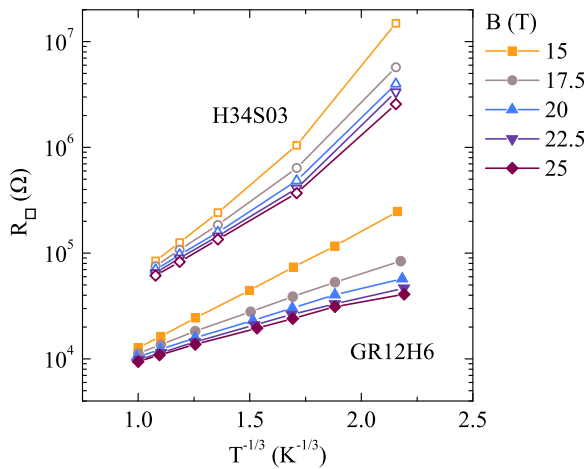} 
\caption{\label{Fig3} Sheet resistance $ R_{\square} $ as a function of $T^{-1/3}$ of our highest disordered superconducting film GR12H6 and the insulating film H34S03 at different magnetic fields, ranging from 15 to 25 T. The data points are extracted from the $ R_{\square} $ versus $ B $ measurements shown in Fig. 1.}
\end{figure}

The study of our samples reveals a different picture. In Figure \ref{Fig3} we show the high-$B$, low $T$ magnetoresistance of samples GR12H6 and H34S03 exhibiting a clear insulating behavior. We found that the $T$-dependence in this regime is best-described by 2D single-electron Mott hopping \cite{efros2}, $R\sim \exp(T_{2D}/T)^{1/3}$, where $T_{2D}$ is the model characteristic $T$ \footnote{While the limited range of our data allows for different $T$-dependences such as 3D Mott hopping ($R\sim \exp(T_{3D}/T)^{1/4}$) or Efros-Shklovskii (ES) hopping ($R\sim \exp(T_{ES}/T)^{1/2}$), where $T_{3D}$ and $T_{ES}$ are the relevant model characteristic $T$, the parameters we obtain from fitting our data to these other forms are inconsistent: by using the 3D hopping form we obtain the localization length, $\xi_L = 60-500$ nm, larger than the film thickness, contradicting the 3D hopping assumption. Similarly, for ES hopping, we obtain $\xi_L\sim 50-1000 ~\mu$m, which is physically unreasonable. For the 2D case that fits best our data we obtain $\xi_L\sim 200-1000\, $nm for sample GR12H6 and $\xi_L\sim 100-200\, $nm for sample H28S03, consistent with a 2D hopping regime}. Sample H34S03 shows deviation from the Mott law for the lowest T points. We did not find any hopping exponent that fits the entire T range including these deviations. This upward deviation is nevertheless consistent with an insulating ground state in the zero temperature limit. With our ability to span a large B-range and achieve very low T's we have thus ruled out metallic behavior in our samples as the phase terminating the CPI at very high $B$. 

Another question raised by our findings pertains to the microscopic mechanism by which the system undergoes the $B$-driven transition from the insulating-peak regime, the CPI where transport is by Cooper-pairs, into the high-$B$ insulating state that, being above $B_\textrm{c2}$, should only comprise unpaired electrons. The observation of such a transition opens the possibility for a new transition, between bosonic and fermionic insulators,  driven by an increase in $B$. The details of this transition awaits further theoretical and experimental inputs.

In summary, we have conducted a comprehensive study of a:InO films spanning the largest $B$-range and the widest range of disorder levels to date. The breadth of our study enabled us to make two new observations: First, we showed that the insulating peak observed in higher disorder samples terminates at a typical $B=13-15$ T that coincides with $B_\textrm{c2}$ in our cleanest samples, providing further evidence for localized Cooper pairing in the insulating peak. Second, we demonstrated that the high-$B$ phase terminating the CPI in a:InO is another insulating phase, having transport properties consistent with conventional Mott hopping of single electrons. The nature of the transition between these two distinct insulators is still unclear.

\begin{acknowledgments}
We are grateful to B. Altshuler, H. Courtois, M. Feigel'man, Th. Grenet, V. Kravtsov, A. Rogachev, and C. Winkelmann for fruitful discussions. This research is supported by the Minerva foundation with funding from the Federal German Ministry for Education and Research, and by the European Community under Contract No. EC-EuroMagNetII-228043.
\end{acknowledgments}

\bibliography{Termination_CPI}
\end{document}